\documentstyle[11pt,newpasp,twoside,epsf]{article}
\def\edcomment#1{\iffalse\marginpar{\raggedright\sl#1\/}\else\relax\fi}
\reversemarginpar

\begin{document}
\title{Outflow from YSO and Angular Momentum Transfer}
 \author{Kohji Tomisaka}
\affil{National Astronomical Observatory, Mitaka, Tokyo 181-8588, Japan}

\begin{abstract}
Dynamical contraction of a slowly-rotating magnetized cloud is studied
 using the magnetohydrodynamical (MHD) simulations.
In the isothermal stage ($n \la n_{\rm A} \sim 10^{10}{\rm cm}^{-3}$),
 the cloud evolves similarly to that expected from
 Larson-Penston self-similar solution and experiences the run-away collapse.
However, after the central density exceeds $\sim n_{\rm A}$,
 an accretion disk is formed around the adiabatic core. 
Just outside the core, outflow is ejected by the effect of 
 magnetic torque (magnetocentrifugal wind).
Since $\sim$ 10\% of the mass is ejected with almost all the 
 angular momentum, the specific angular momentum of the protostellar core
 reduces to that observed in main-sequence stars.  
\end{abstract}

\vspace*{-5mm}\section{Introduction}

Fragmentation of the clouds (cloud cores) is thought to be affected by the 
 amount of angular momentum in it.
On the angular momentum, there has been a problem that
 the specific angular momentum of new born stars (e.g. T Tauri stars)
 $j_* \simeq 6\times 10^{16}(R_*/2R_\odot)^2(P/10{\rm day})^{-1}
 {\rm cm^2~s}^{-1}$ is much smaller than that of the parent molecular cloud
 cores $j_{\rm cl} \simeq 5\times 10^{21}(R/0.1{\rm pc})^2
 (\Omega/4{\rm km\,s^{-1}\,pc^{-1}}){\rm cm^2~s}^{-1}$.
Here we assumed the observed velocity gradient comes from the rotational
 motion.
Further, the specific orbital angular momentum of a binary whose separation
 is about $R\sim 100$AU is equal to
 $j_{\rm orbit}\simeq 4\times 10^{19}(R/100{\rm AU})^{1/2}(M/1M_\odot)^{1/2}$
 ${\rm cm^2~s}^{-1}$.
If the angular momentum is conserved in the star forming process,
 neither binary nor single stars are formed.

Angular momentum is reduced by the effect of magnetic fields (B-fields).
 Considering an oblate spheroidal cloud is threaded by poloidal 
 B-fields parallel to the symmetry axis, the rotation motion generates
 torsional Alfv\'{e}n waves in the inlercloud medium,
 which propagate away from the cloud.
The torsional Alfv\'{e}n wave causes the intercloud medium to rotate.
In this case, the angular momentum in the cloud decreases with a timescale
 as 
$
t_B\equiv -\frac{\Omega}{\dot{\Omega}}\simeq \frac{\sigma}{2\rho_aV_A}
=\frac{2\pi G \sigma}{B_0}(4\pi G\rho_a)^{-1/2},  
$
where $\Omega$, $\sigma$, $\rho_a$, $V_A$, and $B_0$ represent, respectively,
 the angular rotation speed, the column density of the cloud,
 the ambient density, the Alfv\'{e}n speed in the ambient medium, and the 
 magnetic flux density outside the cloud (Ebert et al. 1960).
Since the first factor $2\pi G \sigma/B_0$ is larger than unity for
 magnetically supercritical clouds,
 the timescale $t_B$ is comparable with the free-fall timescale
 in the {\em low-density} ambient medium
 and thus the angular momentum transfer may seem inefficient as long as
 the cloud contracts dynamically.

Molecular outflow begins at some epoch in the contraction stage of the cloud.
The molecular outflow ejects matter and thus angular momentum.
Consider the situation that
 90 \% of the inflowing matter forms a star and the rest 10 \% is
 ejected by the molecular outflow. 
On the other hand, 0.1\% of the angular momentum goes to the newly formed
 star and the rest is ejected by the outflow.
If this occurs, the specific angular momentum of a newly formed star 
 is reduced by a factor $10^{-3}$ from that of the parent cloud.
To confirm whether this mechanism works in a process of the star formation,
 we have studied the evolution of a slowly rotating cloud and
 the formation of molecular outflows.

\section{Numerical Method \& Model}
  
We assume the gas is well coupled with the B-fields.
The basic equations to solve are the ideal MHD equations with cylindrical
 symmetry and the Poisson equation for the gravitational potential.
Since the gravitational collapse proceeds heavily nonhomologously,
 we have to cover the wide dynamic ranges for both the density and 
 the size.
This was attained by using the nested grid method, in which
 a number of grids with different spatial resolution are prepared
 (here we used 24 grids from the coarsest L0 to the finest L23)
 and calculated simultaneously.
The finer grids are for small-scale structure near the core
 while the coarser ones cover the global structure.

As the initial condition, we assume a cylindrical isothermal cloud in 
 hydrostatic balance.
B-field strength is taken as proportional to $\rho^{1/2}$ and the cloud's
 rotation axis coincides with the axis of the cylinder. 
Density perturbations with $\lambda$ equal to that of the most unstable
 mode are added.

\section{Run-away Collapse Stage}

Isothermal clouds experience the run-away collapse, in which
 the central density increases very rapidly in a finite timescale.
This is common for spherical clouds without rotation and
 B-fields (Larson 1969) and for a disk formed in a rotating isothermal
 cloud (Norman et al 1980).
The gas falling down along the B-fields forms a contracting
 disk (pseudo-disk) even for rotating and magnetized clouds. 
Contraction continues if the equation of state remains isothermal.
In this phase, there is no evidence for the outflow (Fig.1a,c).

\section{Accretion Stage}

After the central density reaches $\rho_{\rm A}\sim 10^{10}{\rm cm}^{-3}$,
 the thermal radiation from dust grains is trapped in the central core and
 an adiabatic core is formed (Larson 1969).
Since the dynamical timescale of the core becomes much shorter than
 that of the outer isothermal envelope, the isothermal gas begins to
 accrete to the nearly hydrostatic adiabatic atomic core (first core). 
To study the evolution of this accretion stage,
 we use a composite polytrope model as
 below $\rho_{\rm A}$ the gas is assumed isothermal;
 for  $\rho > \rho_{\rm A}$  the gas becomes adiabatic and the polytropic
 index $\Gamma$ is assumed as 5/3 (see Tomisaka 1998).

Figure 1 illustrates the change from the isothermal run-away collapse (Fig1.a, c)
 to the adiabatic accretion phase (Fig.1b, d).
Figure 1a shows that B-fields (red lines) run almost
 perpendicularly to the disk (blue surfaces) in the run-away collapse phase.
Toroidal B-fields and rotational motion are not predominant (Fig.1c).
On the other hand, in the accretion phase B-fields are squeezed by the
 effect of inflow motion in the disk and they form an hourglass-shape
 configuration.
Further, toroidal B-fields are formed by the rotational motions which
 are amplified in the accretion phase (Fig.1d).
In this phase, outflow occurs (Fig.1b, d) 
 at the distance $\simeq 50 - 100$ AU from the center.
The time elapsed between these two is only about 1000 yr.

The Lorentz force ${\bf F}=(c/4\pi){\bf j}\times {\bf B}$ contains the toroidal
 component $F_\phi=(c/4\pi){\bf j_{\rm pol}}\times {\bf B}_{\rm pol}$ when
 there exists the poloidal current (${\bf j_{\rm pol}}$)
 which makes the toroidal B-fields $B_\phi$.
Therefore, it is shown that 
 the magnetic torque $N=r\, F_\phi$, which changes the
 distribution of the angular momentum, works effectively 
 only after the toroidal B-field is amplified by the rotation motion.
The toroidal B-fields become predominant after the core formation epoch.
Further, Blandford \& Peyne (1982) showed that efficient angular
 momentum transfer from the disk to the (cold) wind material occurs
 only when the angle between the B-field line and the disk is smaller
 than 60\deg.
This condition is satisfied only in the accretion phase (compare Fig.1a and.1b)

\section{Angular Momentum Transfer}

\begin{minipage}[b]{0.5\textwidth}
To study the effect of angular momentum transfer by the magnetic torque,
 specific angular momentum $j$ is plotted against the accumulated
 mass from the center $M$ in Figure 2 (Tomisaka 2000).
The difference between the initial stage (dashed line)
 and the core formation epoch (dotted line)
 does not come from the angular momentum transfer but from the segregation
 of the mass with small $j$ to form the high-density central part.
True transfer occurs after the core formation.
The magnetic torque increases $j$ of the matter on the disk surface and
 the matter with excess $j$ is ejected as the outflow,
 while the magnetic torque decreases $j$ inside the disk and this leads to 
 accelerate the accretion.

\hspace*{8mm}That is, B-fields changes the angular momentum distribution along one
 field line. 
As a result, in $\simeq$7000 yr after
\end{minipage}
\hspace*{1mm}
\begin{minipage}[b]{0.48\textwidth}
{\plotone{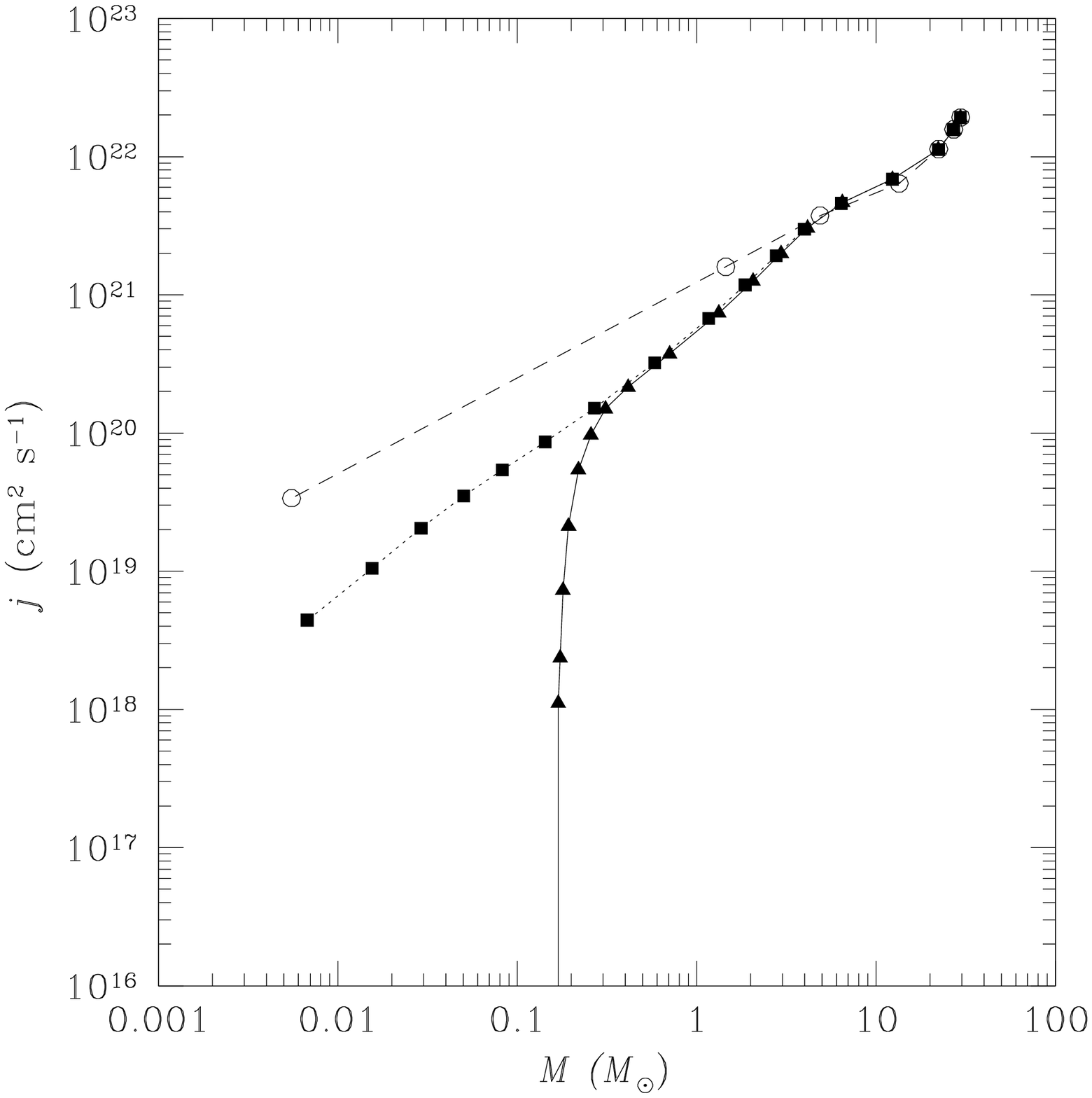}}\\
{\small Figure 2:
Specific angular momentum $j$ is plotted against the accumulated
 mass from the center $M$.  
Open circles, filled squares, and triangles denote, respectively,
 the beginning and the end of the run-away collapse phase and the 
 accretion phase.
} 
\end{minipage}
 the core formation $M\simeq 0.2M_\odot$
 is accumulated in the core and $j$ of the core becomes as small as 
 $\sim 10^{16}{\rm cm^2\, s}^{-1}$ (solid line).
Therefore, it is clearly shown that the angular momentum problem 
 of newborn stars is solved taking into account the molecular outflow
 driven by the magnetic torque formed just outside the first core
 ($R\la 100 $AU).

\begin{figure}[h]
\hfill {\small (a) \hfill (b)} \hfill \\
\plottwo{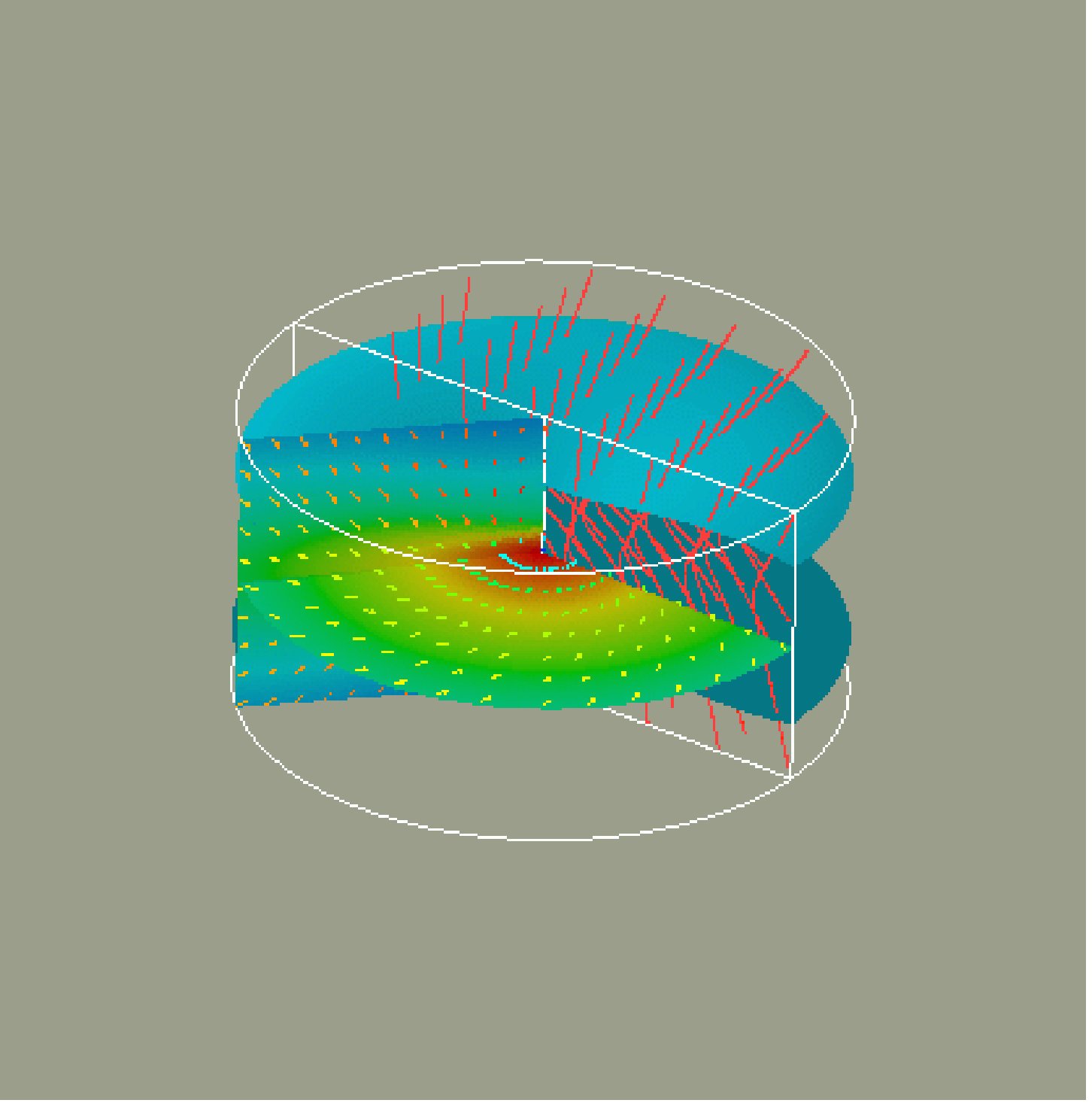}{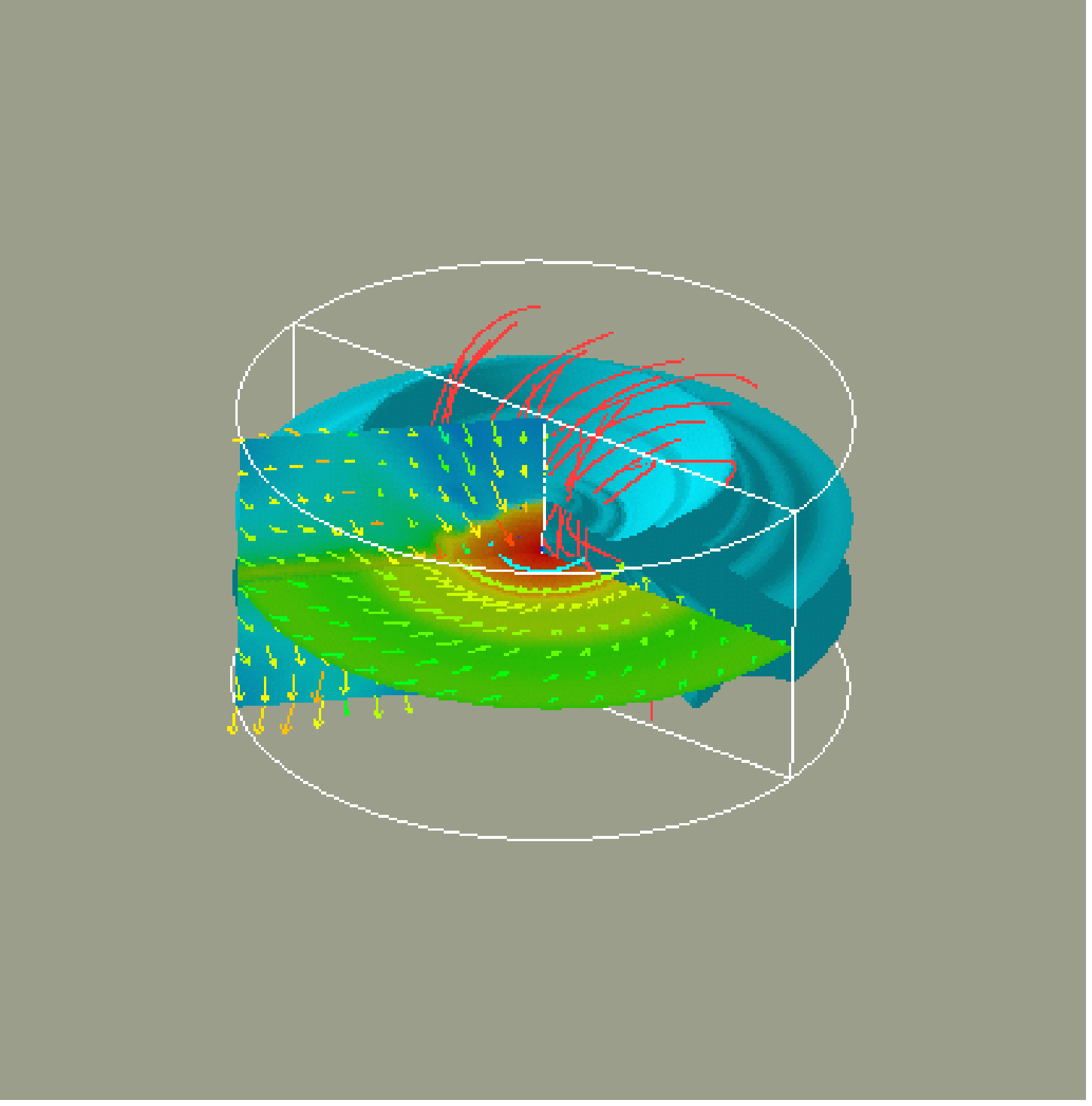}\\
\hfill {\small (c) \hfill (d)} \hfill \\
\plottwo{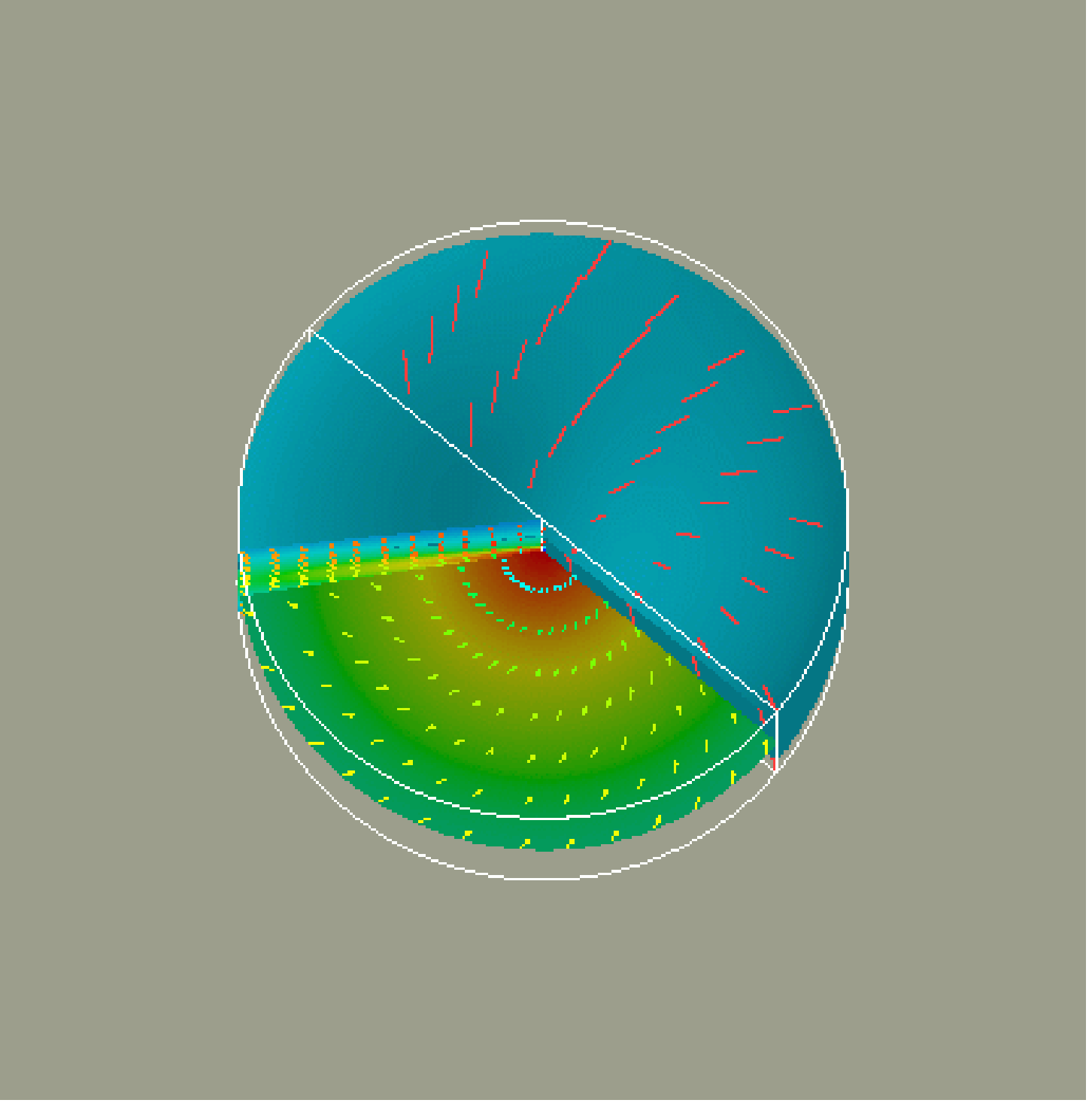}{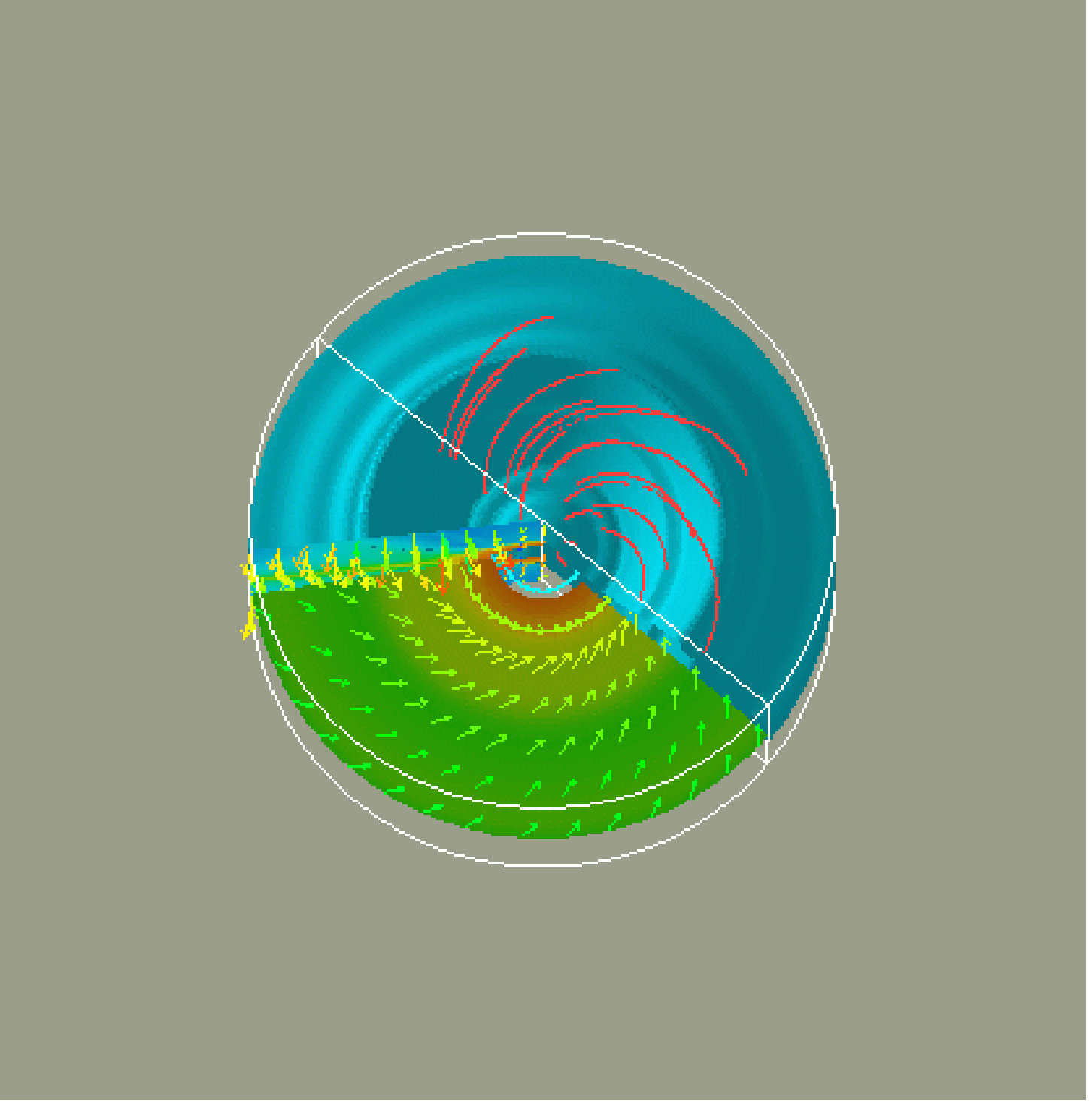}\vspace*{-5mm}
\caption{
The left panels illustrate the structure in the run-away phase,
 while the right ones are for the accretion phase.
The upper panels represent nearly edge-on views and
 the lower ones are pole-on views.
These represent the structure captured by L10 grid and the diameter
 of the numerical box (white circle) is equal to $\simeq$ 300 AU.
Isodensity surfaces and B-field lines are illustrated as well as
 the velocity vectors and density countours.
}
\end{figure}

\end{document}